\documentclass[aps,
nofootinbib,
12pt
]{revtex4}

\newcommand{\beq}{\begin{equation}}
\newcommand{\eeq}{\end{equation}}
\newcommand{\ber}{\begin{eqnarray}}
\newcommand{\eer}{\end{eqnarray}}

\newcommand{\eeql}[1]{\label{#1}\eeq}

\input epsf

\newcommand{\figin}[2]{
\begin{figure}[t]
\centerline{\hbox{\epsffile{#1.eps}}}
\centerline{\parbox{8cm}{\caption{#2\label{#1}}}}
\end{figure}}

\newcommand{\fig}[1]{Fig.~\ref{#1}.}
\newcommand{\figur}[2]{\fig{#1}\figin{#1}{#2}}

\newcommand{\vev}[1]{{\left< {#1} \right>}}

\newcommand{\half}{{1\over 2}}

\begin{document}

\begin{table}[t]
\begin{flushright}
   WIS/28/02-July-DPP\\
   hep-th/0207266
\end{flushright}
\end{table}

\title{Closed String Amplitudes from Gauge Fixed String Field Theory}
\author{Nadav Drukker}
\affiliation{Department of Particle Physics, Weizmann Institute of Science, 
Rehovot 76100 Israel}
\email{drukker@weizmann.ac.il}
\date{July 30, 2002}
\begin{abstract}

Closed string diagrams are derived from cubic open string 
field theory using a gauge fixed kinetic operator. The basic idea 
is to use a string propagator that does not generate a boundary 
to the world sheet. Using this propagator and the closed string 
vertex, the moduli space of closed string surfaces is covered, so 
closed string scattering amplitudes should be reproduced. This kinetic 
operator could be a gauge fixed form of the string field theory 
action around the closed string vacuum.

\end{abstract}

\maketitle

\vfill\eject
\section{Introduction}

Over the past few years there has been a resurgence of interest in 
string field theory as a tool in studying non-perturbative effects 
in string theory. It was proposed by Sen \cite{Sen:1999mh} 
that condensing the 
open string tachyon in bosonic string theory will lead to the 
annihilation of the brane and to the closed string vacuum. In 
string field theory this should be described by a classical solution 
whose action is minus the tension of the annihilated brane.

This classical solution was studied numerically using level 
truncation techniques \cite{Kostelecky:1989nt,Sen:1999nx,Moeller:2000xv} 
in Witten's cubic string field theory \cite{Witten:1985cc}, but to 
date the analytical solution was not found. 
Given a classical solution one can re-expand the action around 
the new vacuum which will result in a new kinetic term, but the 
same cubic interaction vertex. Instead of finding the classical 
solution one can guess a form for the kinetic operator in the closed 
string vacuum and start from there.

Gaiotto et al. \cite{Gaiotto:2001ji} proposed a pure ghost midpoint 
insertion as the new kinetic operator. That led to some beautiful 
results, but is rather singular. Therefore it makes sense to look 
for other forms of the kinetic operator in the closed string vacuum.

As a guide to finding this kinetic operator we look at the Feynman 
rules and find kinetic operators 
that reproduce closed string diagrams. The Feynman rules were 
derived from the singular ghost insertion in 
Ref.~\onlinecite{Gaiotto:2001ji} using some regularization. The 
propagator we propose here will be less singular and will not 
suffer some of the problems encountered there.

In Feynman-Siegel gauge the kinetic operator for the open string 
is $c_0(L_0-1)$, where $c_0$ is the ghost zero mode and $L_0-1$ 
the open string Hamiltonian. The propagator is
\beq
G=g_s^2 b_0\int dt\, e^{-t(L_0-1)}\,,
\eeq
where $b_0$ is the antighost zero mode. The expression $\exp -t(L_0-1)$ 
can be visualized as generating an open worldsheet of length $t$. The 
variables $t$, which are integrated over, become the Feynman 
parameters of the string graphs and parameterize the moduli space of 
Riemann surfaces.

This propagator is clearly unsuited for closed strings, since it 
adds boundaries to the world sheet. If instead of $L_0$ one 
uses\footnote{
In Ref.~\onlinecite{Drukker:2003hh} it is shown how similar expressions 
arise from the gauge invariant kinetic terms of 
Ref.~\onlinecite{Takahashi:2002ez}, whose cohomology is trivial
\cite{Kishimoto:2002xi}. Using that as the starting point one finds 
that apart for the 
usual energy momentum tensor there is an extra ghost contribution 
appropriate for the twisted $b,c$ system with central charge $-2$. 
}
\beq
\check L_0={1\over\pi}\int d\sigma \sin\sigma(T+\bar T)\,,
\eeql{newL}
this operator will generate a worldsheet without adding any boundary. 
The reason is that the energy momentum tensor will not act at 
$\sigma=0$ and at $\sigma=\pi$. This is a realization of an idea 
presented in Ref.~\onlinecite{Shatashvili:2001ux} that closed strings 
should arise from boundaries shrinking to points.

In the next section the Feynman rules are derived using this kinetic 
operator and it is demonstrated how they lead to all closed string 
diagrams.

In section 3 we propose a larger family of kinetic 
operators, which are all appropriate for generating closed surfaces. 
They are all related to each other by conformal transformations. In 
particular, we recover the singular propagator of 
Ref.~\onlinecite{Gaiotto:2001ji}.

We end with some comments.

\section{Feynman rules}

We wish to derive the Feynman rules from the cubic string field theory 
action with the kinetic operator replaced by (\ref{newL}). 
The action is
\beq
S={1\over g_s^2}\int
\half\Phi\star c_0 \check L_0 \Phi
+{1\over 3}\Phi\star\Phi\star\Phi\,.
\eeq
Here $\Phi$ is a string field, and the star and integral are defined 
as usual on string fields \cite{Witten:1985cc}. The only difference 
compared to the standard gauge fixed form is the kinetic operator, 
involving $c_0$ and $\check L_0$, whose mode expansion is
\beq
\check L_0={2\over\pi}\sum_{n=-\infty}^\infty
{1\over 1-4n^2}L_{2n}\,.
\eeql{losc}
To avoid having too many fields in the theory, one must impose a gauge 
condition of the fields. The appropriate one seems to be 
$b_0 \Phi=0$.

The propagator derived from this action is
\beq
G=g_s^2 b_0\int dt\, e^{-t\check L_0}\,.
\eeq
Ignoring the $b$ insertion 
for now, the propagator for given $t$ generates a worldsheet that is a 
segment of a sphere. The points $\sigma=0,\pi$ are the north and south 
pole, and are not moved by this propagator. The equator at 
$\sigma=\pi/2$ is moved forward a distance $t$.

Closed strings do not exist as external states in open string field 
theory. Instead, one can use the open-to-closed string vertices of 
\cite{Shapiro:ac,Shapiro:gq,Zwiebach:1992bw}, which can be regarded as 
the gauge invariant observables of the theory 
\cite{Hashimoto:2001sm,Gaiotto:2001ji}. For any closed 
string vertex operator $V=c\bar c V_m$, where $V_m$ is a dimension 
$(1,1)$ primary in the matter CFT, one defines the gauge invariant 
operator
\beq
O_V=\int V(\textstyle{\pi\over2})\Phi\,.
\eeq
In Feynman diagrams the closed string vertex $V$ will be inserted at 
the midpoint at the end of a propagator and the right and left parts 
of the string will be identified by the integral.

Closed string amplitudes are the correlators of any number of those 
gauge invariant operators. We will concentrate on the three-point 
function
\beq
\vev{O_1 O_2 O_3}\,.
\eeql{OOO}
At tree level it is given by one cubic interaction vertex, three 
propagators extending from it, with a closed string vertex at the end 
of each. This is depicted schematically in \figur{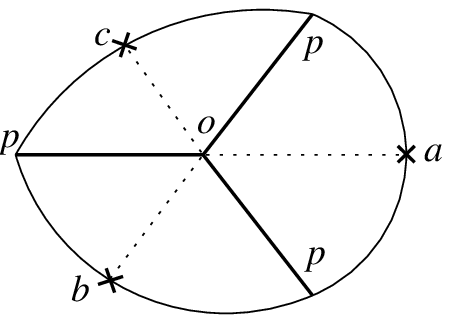}
{A schematic depiction of the tree level Feynman diagram contributing 
to the three point function of closed string vertices. The three lines 
$op$ make up the interaction vertex, and attached to them are three 
propagators. At $a$, $b$ and $c$ closed string vertex operators are 
inserted, and the two halves of the strings labeled by $pa$, by $pb$ 
and by $pc$ are identified with each other.}

To better visualize the worldsheet it is useful to cut the three 
propagators along their midpoints. Since the propagators were segments 
of a sphere, we find now six segments of a hemisphere glued to each 
other, forming one segment of the hemisphere. If the lengths of the 
three propagators are $t_1,t_2,t_3$, the total angle around the 
hemisphere is $2(t_1+t_2+t_3)$. Since the two ends of the segment are 
identified, there could be a conical singularity at the pole. This is 
shown in \figur{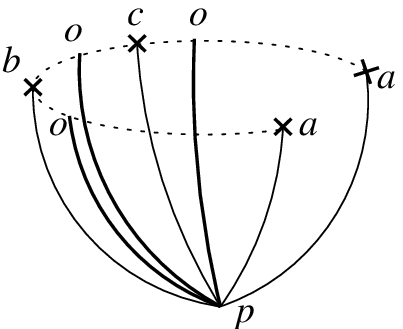}
{Another depiction of the same diagram as in Fig.~1. Here we cut the 
propagators along the midpoints $oa$, $ob$ and $oc$, and glued the 
ends of the propagators marked by $pbp$ and $pcp$. The result is 
a segment of the southern hemisphere, with a deficit angle (where 
we should glue the two lines marked $ap$).}

If we fix the total length of the three propagators $t=t_1+t_2+t_3$, 
the integration over the relative lengths covers the moduli space of 
spheres with four marked points (the three vertices and the pole). 
This fact will be proven in the next subsection. The three $b$ 
insertions can be represented as contour integrals on the Riemann 
surface. Two of them should be associated with the two moduli giving 
the correct measure over the moduli space. The third, associated 
with the integration over the total angle, should give some measure 
for that integral.

So at tree level the correlator of the three closed string vertices 
in string field theory (\ref{OOO}) will be given by a conformal field 
theory calculation on the sphere with the closed string vertices and 
one extra marked point. This marked 
point is the remnant of the shrunken boundary, and we should take 
care to see what happens there. If, for example, we started with all 
Neumann conditions we will find that no momentum flows through this 
marked point. This looks like a zero momentum vertex insertion, and 
may be a soft dilaton as proposed in \cite{Gaiotto:2001ji}.

Instead, if we had Dirichlet 
boundary conditions on the open strings, we will be left with 
the constraint that this point on the worldsheet is mapped to a 
fixed point in space. To remove this constraint we have to integrate 
over all locations of this D-instanton. Alternatively, one may impose 
periodic boundary conditions~\cite{Moore:2001fg}, 
which will not restrict the position of 
this point, or the momentum flowing through it, but this is not very 
natural for open strings.

The result of the conformal field theory calculation is multiplied by 
a power of the open string coupling $g_s^4$ and by the integral over 
the extra parameter $t$, call it $T$ (which possibly diverges). 
At higher order in perturbation theory there are 
graphs with more marked points. Those come from open string diagrams 
with more than one boundary, each shrinking to a point. The graph 
with $m$ shrunken boundaries will have a factor of $g_s^{2+2m}$, and 
the angle around each marked point is still arbitrary, giving 
something like $T^m$. Therefore we get the final result
\beq
\vev{O_1 O_2 O_3}
\sim g_s^2\sum_m (g_s^2 T)^m \vev{V_1 V_2 V_3}_{{\rm sphere},m}\,,
\eeql{sum}
where the last correlation function is the conformal field theory 
result for the sphere with three vertex operators and the $m$ marked 
points (treated in one of the ways proposed above, or differently).

Other graphs will give similar expressions for the torus and higher 
genus contributions to this amplitudes. The structure will be very 
similar, with a series in $g_s^2 T$ multiplying the correct power of 
the coupling. The same is true for correlators of more closed string 
vertices.

\subsection{Covering of moduli}

To complete the discussion on the closed string diagrams it is 
necessary to prove that integrating over the Feynman parameters in 
the string diagrams covers the moduli space of Riemann surfaces. There 
are two standard ways of demonstrating this, one is by checking 
that the limits of integration do not leave holes in the moduli space, 
so that all corners of moduli space can be recovered from the Feynman 
parameters.

The other method is to find a minimal area problem that leads to the 
same surfaces as the string diagrams. For every point in moduli space 
there will be a minimal surface that will correspond to some string 
diagram \cite{Giddings:1986wp,Zwiebach:1990az,Zwiebach:1990ba}.

One can use the first method directly. If we represent the surface 
like in Fig.~2. as a hemisphere with a conical singularity at the pole. 
This is similar to the open string diagram presented in Fig.~4 of 
Ref.~\onlinecite{Giddings:1986wp}, where instead of a hemisphere 
there is a cylinder. There too one has to glue segments of the top of 
the cylinder, and the moduli space is the same, except for the modulus 
associated with the size of the boundary, which now is the fixed angle 
$2t$. The other two parameters represent the relative size of 
the six segments at the equator (they are pairwise equal). The corners 
of moduli space are when two vertex operators approach each other, 
which clearly can be done.

Like in the open string case, the same is true for arbitrary diagrams. 
At higher loops one can get disconnected diagrams with several 
hemispheres, and one has to sum over all different ways of gluing a 
fixed number of segments on them.

Instead of making this argument more rigorous, we can use the second 
method of minimal surfaces. This technique cannot be applied directly, 
since minimal surfaces have to be hyperbolic, or flat (except for 
singular points).
Our surfaces have constant positive curvature, 
apart for the singular points. The trick is to map the hemisphere to a 
semi-infinite cylinder using $u=\ln\tan(\sigma/2)$. This is a 
conformal transformation, so we can study all the surfaces with this 
flat metric, instead of the curved one. This can be seen in 
\figur{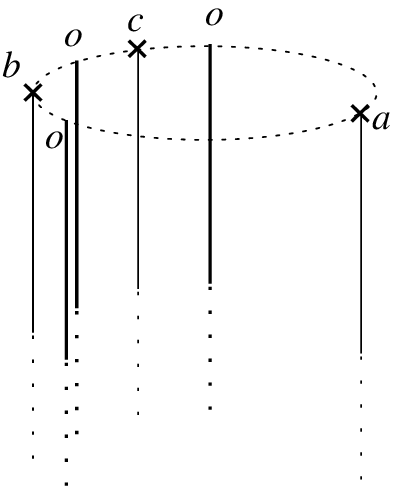}
{By a conformal map it is possible to map the hemisphere in Fig.~2. 
to a semi-infinite cylinder of circumference $2t$. The point $p$, 
which is the remnant of the boundary was pushed to infinity.}

The original surfaces had in general $n$ closed string punctures, and 
$m$ shrunken boundaries with angles $2t_i$. The new surfaces will be 
made up of $m$ semi-infinite tubes of circumferences $2t_i$ 
and with $n$ punctures where they are glued.
Those surfaces solve a minimal area problem very similar to the one 
discussed in Ref.~\onlinecite{Gaiotto:2001ji}, and the proof carries 
over.

Consider a Riemann surface with $n$ punctures where vertex 
operators are inserted and $m$ more punctures, for $m\geq1$, and 
positive numbers $t_i$ with $i=1,\cdots,m$ associated to them.
Find the minimal surface such that all curves homotopic to the 
puncture $i$ are at least of length $2t_i$. If one does not want to 
distinguish between the two sets of punctures, one just assigns 
$t=0$ to the punctures where the vertex operators are inserted.

It is not hard to find the solution to this problem. For any puncture 
at a finite distance on the surface there are arbitrarily small 
curves around it. Since all curves homotopic to the puncture $i$ have 
a length $2t_i>0$, the puncture cannot be at a finite distance on the 
surface. So it sits at the end of a semi-infinite cylinder.

The surface is therefore comprised of $m$ semi-infinite cylinders glued 
to each other. The $n$ other punctures are inserted along the line 
where the cylinders are glued. One should note that those surfaces are 
minimal in the sense that they solve the Euler Lagrange equation with 
the constraint. The area is infinite, so one cannot calculate the area 
of different surfaces and compare without some regularization. 
Still all the vertex operators should be along the same 
circles where the cylinders are identified. Otherwise there would be 
a finite piece of the cylinder that could be removed and the vertex 
operators brought closer.

By this method one can get any Riemann surface of any genus, as long 
as $m\geq1$. The main example we studied before was a sphere with $n=3$ 
and $m=1$. Another example is the torus, and to keep it simple we take 
$n=0$ and $m=1$, so it represents the one-loop vacuum amplitude. Since 
$m=1$ this would be the disc with one handle if we used the regular 
propagator of open string field theory. That surface was constructed in 
Ref.~\onlinecite{Giddings:1986wp}. It is made up of two three-point 
vertices connected by three propagators. One can connect the propagators 
in different ways, either giving a planar surface with three boundaries, 
or a surface with a single boundary, but a handle. We are interested 
in the second case, where $m=1$.

Using the new propagator and slicing each propagator along the midpoint 
one ends up with the same geometry as depicted in Fig.~2. Again, six 
segments along the equator have to be identified pairwise. But before, 
there were closed string vertices along the equator and the two 
segments around each of the vertices were identified (so we could label 
them as $112233$). Now there are no closed string vertices, and neighboring 
segments should not be identified, so as to get the non-trivial 
topology of the torus (the arrangement is $123123$).

By the conformal map, this is again a semi-infinite cylinder with the 
identifications at the end, and this is a solution of the minimal area 
problem presented above for the torus with $m=1$ and $n=0$. This surface 
has two 
moduli, the ratio of the lengths of the segments that are identified, 
and the torus degenerates as the ratio goes to zero. The total 
circumference is the parameter $t$ as before, which is integrated 
over, but is not a modulus for the Riemann surface, since it corresponds 
to a scaling. Indeed a torus with one marked point has two real moduli.

\section{Generalizations}

The  kinetic operator $b_0 \check L_0$ is not unique in 
generating closed 
surfaces. Instead of the function $\sin\sigma$ appearing in the 
definition (\ref{newL}) we can use any other function $f(\sigma)$ 
subject to the constraints that it is positive, except for a simple 
zero at $f(0)=f(\pi)=0$, and that $f(\pi-\sigma)=f(\sigma)$. The 
first condition is required so the propagator will not generate a 
boundary and the second one so gluing the two halves of the string 
would not generate a line singularity.

The factor of $\sin\sigma$ in the definition of $\check L_0$ 
corresponds to the standard metric on the sphere 
$ds^2=d\sigma^2+\sin^2\sigma dt^2$. 
For a general function the propagator will generate segments of a 
deformed sphere, with the metric $ds^2=d\sigma^2+f(\sigma)^2 dt^2$.

One special case is to take $f(\sigma)=\lim_{\epsilon\to 0}\epsilon$, 
which gives the sphere squeezed into a very thin cylinder. This is 
very close to the regularized propagator considered in 
Ref.~\onlinecite{Gaiotto:2001ji}. To be more rigorous, one should take a 
function that is equal to $\epsilon$ almost everywhere, and 
approaches zero at the boundaries. If one rescales the infinitesimal 
cylinder to finite size it will give the Feynman graphs considered at 
the end of the last section, and depicted in Fig.~3.

It is easy to prove the equivalence of all those propagators, a simple 
coordinate transformation maps the sphere with the funny metric to the 
usual sphere
\beq
\ln\tan{\sigma\over 2}=\int{d\sigma'\over f(\sigma')}\,.
\eeq
This is a conformal transformation, so all those graphs are equivalent. 
Clearly the two simplest choices are the round sphere and the infinite 
cylinder.

\section{Conclusions}

We have constructed a family of gauge fixed kinetic operators on open 
string fields that give closed string diagrams when used to calculate 
the correlators of closed string vertices.

This construction could serve as a guide to finding the gauge 
invariant kinetic operator at the vacuum of string field theory. This 
would be analogous to the way the BRST operator was identified as the 
gauge invariant kinetic operator for open string field theory 
\cite{Siegel:1984xd,Siegel:1985tw,Itoh:1985bb,Banks:1985ff}.

It would be interesting to check these operators in a real 
calculation. One way to do that is doing the explicit conformal field 
theory calculations, mapping the surfaces to the sphere like was done 
for open strings in Ref.~\onlinecite{Giddings:1986iy}. One may also 
use an algebraic approach, like in Ref.~\onlinecite{Taylor:2002bq}, 
but in the standard oscillator basis the kinetic operators that vanish 
on the boundary (\ref{losc}) are very non-diagonal. The calculation 
might be simpler in a different basis.

Another problem is the nature of the series in equation (\ref{sum}). 
This kind of sum should be expected in any formalism of closed 
strings in terms of open strings. Starting with a Riemann surface 
with no handles, but any number of boundaries and shrinking them to 
points will give the sphere with extra marked points. So many open 
string diagrams will lead to the same closed string diagram. In 
particular it would be nice if $T\sim 1/g_s$, and the series 
converged to some number.

This kind of kinetic operator can be used also for the superstring, 
again generating closed surfaces. But finding the interaction terms in 
the action at the closed string vacuum might be a difficult problem.

\section*{Acknowledgments}

I am grateful to Oren Bergman, Sunny Itzhaki, Harlan Robins, 
Adam Schwimmer, Joan Sim\'on, Wati Taylor and Barton Zwiebach 
with whom I discussed this subject and who made very helpful 
suggestions. 
I would also like to thank the organizers of the 
Carg\`ese summer school for providing a great environment that enabled 
me to complete this work.


\end{document}